\documentclass[conference]{IEEEtran}
\IEEEoverridecommandlockouts
\usepackage{cite}
\usepackage{amsmath,amssymb,amsfonts}
\usepackage{subcaption}
\usepackage{array}
\usepackage{hyperref}
\usepackage{tikz}
\usepackage{algorithmic}
\usepackage[para]{footmisc}
\usepackage{graphicx}
\usepackage{balance}
\usepackage{textcomp}
\usepackage{xcolor}
\usepackage[T1]{fontenc}
\def\BibTeX{{\rm B\kern-.05em{\sc i\kern-.025em b}\kern-.08em
    T\kern-.1667em\lower.7ex\hbox{E}\kern-.125emX}}
\begin{document}

\title{The \guillemotleft Huh?\guillemotright{}  Button: Improving Understanding in Educational Videos with Large Language Models}

\author{\IEEEauthorblockN{Boris Ruf}
\IEEEauthorblockA{\textit{AI Research} \\
\textit{AXA Group Operations}\\
Paris, France \\
boris.ruf@axa.com}
\and
\IEEEauthorblockN{Marcin Detyniecki}
\IEEEauthorblockA{\textit{AI Research} \\
\textit{AXA Group Operations}\\
Paris, France \\
marcin.detyniecki@axa.com}
}

\maketitle

\begin{abstract}
We propose a simple way to use large language models (LLMs) in education. Specifically, our method aims to improve individual comprehension by adding a novel feature to online videos. We combine the low threshold for interactivity in digital experiences with the benefits of rephrased and elaborated explanations typical of face-to-face interactions, thereby supporting to close knowledge gaps at scale. To demonstrate the technical feasibility of our approach, we conducted a proof-of-concept experiment and implemented a prototype which is available for testing online. Through the use case, we also show how caching can be applied in LLM-powered applications to reduce their carbon footprint.\end{abstract}


\section{Introduction}

When sharing knowledge, different knowledge backgrounds among individuals can give rise to misunderstandings, hindering the comprehension of new concepts. This phenomenon often results in individuals "losing the thread" of an explanation. In academic settings, such as during lectures, students can interrupt the speaker to seek clarification. This one might choose to reword their explanation or offer supplementary information. However, because of timidity or a reluctance to disrupt the learning experience for others in the audience, a listener may hesitate to do so. Access to recorded lectures presents a potential solution, as it allows for barrier free review of unclear segments. However, simply replaying the same content does not always effectively address the underlying causes of misunderstanding, for example when they are linked to gaps in knowledge or indistinct speech.

In this workshop paper, we propose merging the benefits of the two described scenarios. By leveraging existing data and technologies in an innovative manner, we illustrate how a straightforward feature could enhance learners' comprehension of educational content.

\section{Related work}

Numerous researchers and practitioners are actively exploring new ways to leverage technology in the education sector\cite{NBERw23744}. In particular the emergence of artificial intelligence (AI)  has paved the way for new approaches to support individual learning and training\cite{holmes2022,zhai2020}. One approach is the use of personalized learning systems, which can adapt to the unique needs and abilities of each student. These systems can incorporate a range of educational content, including video lectures, interactive simulations, and online assessments, to deliver a customized learning experience that is tailored to the individual learner's needs\cite{murtaza2022ai,pratama2023revolutionizing}. The researchers are also investigating the use of conversational agents such as chatbots to enable customized tutoring in a personalized teaching experience\cite{hiremath2018chatbot,clarizia2018chatbot}.

However, plenty of barriers to technology adoption in education have been observed~\cite{rogers2000}, and the big educational technology transformation has yet to materialize~\cite{reich2021}. Many "ed tech" products and services are not successful because their conceptual ideas exceed the limits of what is technically possible in production mode~\cite{Mahbub_2023}. Developing a fully interactive learning experience with a customized avatar remains a complicating technical challenge~\cite{vallis2023}, not to mention the time consuming task of producing sophisticated learning content. 

Extensive research has been conducted on the influence of video lectures on the learning process. Best practices for implementation as learning tool while taking into account students' perceptions and learning results have been identified~\cite{fleck2014youtube,duffy2008using}. Several studies have found that students exhibited enhanced comprehension and recall of intricate concepts when exposed to visual explanation videos\cite{6246045,burke2009assessment}. 

Video platforms such as YouTube have been the subject of numerous technical experiments, particularly involving the utilization of transcripts. One study analyzed the improvements in automatic speech recognition systems~\cite{6707758}. Another study proposed a method to automatically detect misinformation in YouTube videos~\cite{10.1145/3340555.3353763}. In a different work, researchers presented a system that uses subtitles to summarize YouTube lecture videos, aiming to improve the efficiency of video content search~\cite{10.1145/3397482.3450722}.

Problems of understanding in educational settings have been systematically explored long before the emergence of video platforms, recognizing the concept of rephrasing as one potential solution\cite{van1985students,kendrick1990problems}.

\section{Our approach}
According to a study conducted by Statista, educational videos are one of the most frequently viewed form of online video content among internet users~\cite{statista}. With the staggering amount of content uploaded to YouTube every minute exceeding 500 hours~\cite{youtube}, it can be inferred that millions of videos featuring educational or tutorial material are accessible on this online video sharing platform. Currently, the prevailing method for enhancing comprehension whilst viewing such content involves the utilization of subtitles, coupled with the rewinding of segments which are unclear to the viewer. Both options offer a distinct advantage over traditional in-person lectures. 

In particular, the lowered inhibition threshold to interrupt the presentation and seek clarification during a digital experience is a notable benefit. A disadvantage of this approach, however, is that an exact verbatim repetition of the spoken content does not necessarily facilitate understanding. Words may have been pronounced unclearly or a formulation may be misleading. As each person possesses a unique background and distinct prior knowledge, individual knowledge gaps may prevent the learner from connecting the dots and comprehend the material. In a live presentation, a speaker could potentially overcome this issue by rephrasing and elaborating upon the content\cite{McNamara2023}. In a recorded lecture, this form of support is not available.

\begin{figure}[t]

\renewcommand{\thefootnote}{\arabic{footnote}}
  \begin{subfigure}[t]{.23\textwidth}
  \centering
  \includegraphics[height=7.3em]{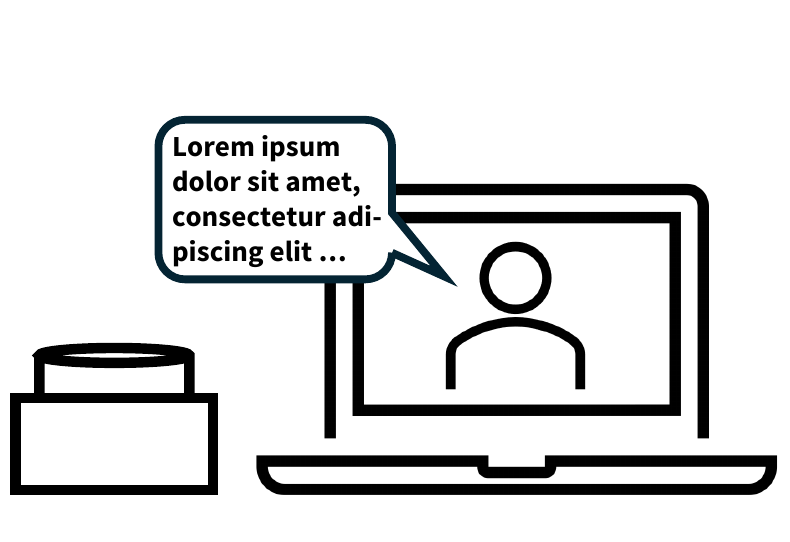}
  \caption{Video lecture streaming}
  \label{fig:sub-first}
\end{subfigure}\quad
\begin{subfigure}[t]{.23\textwidth}
  \centering
  \includegraphics[height=7.3em]{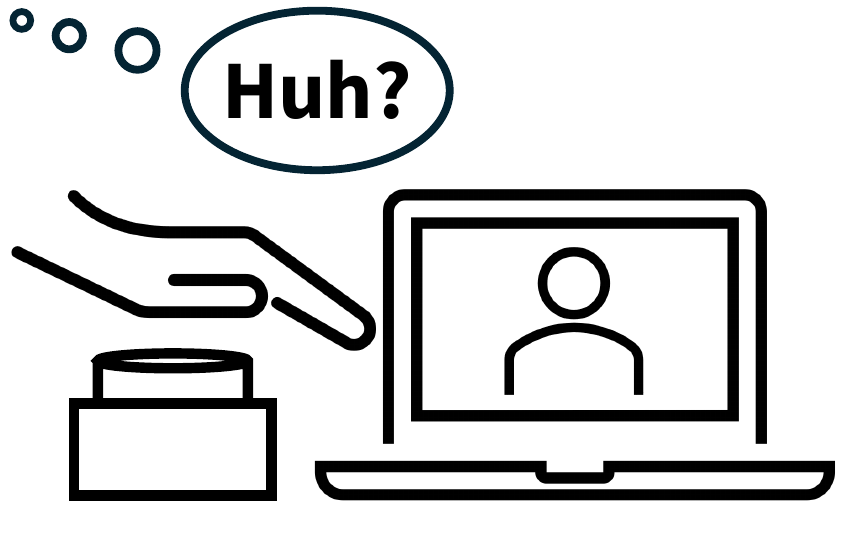}
  \caption{Viewer seeks clarification}
  \label{fig:sub-second}
\end{subfigure}\quad
\begin{subfigure}[t]{.23\textwidth}
  \centering
  \includegraphics[height=7.3em]{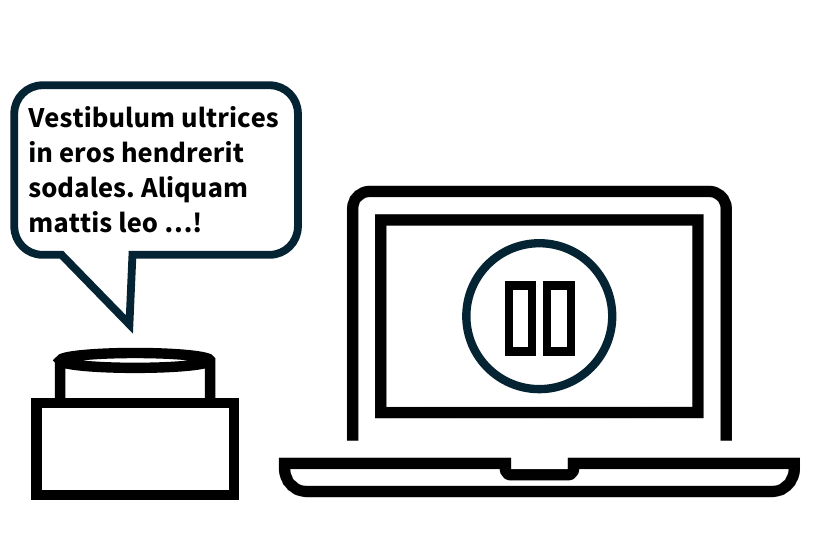}  
  \caption{Video stops, AI explains}
  \label{fig:sub-second}
\end{subfigure}\quad
\begin{subfigure}[t]{.23\textwidth}
  \centering
  \includegraphics[height=7.3em]{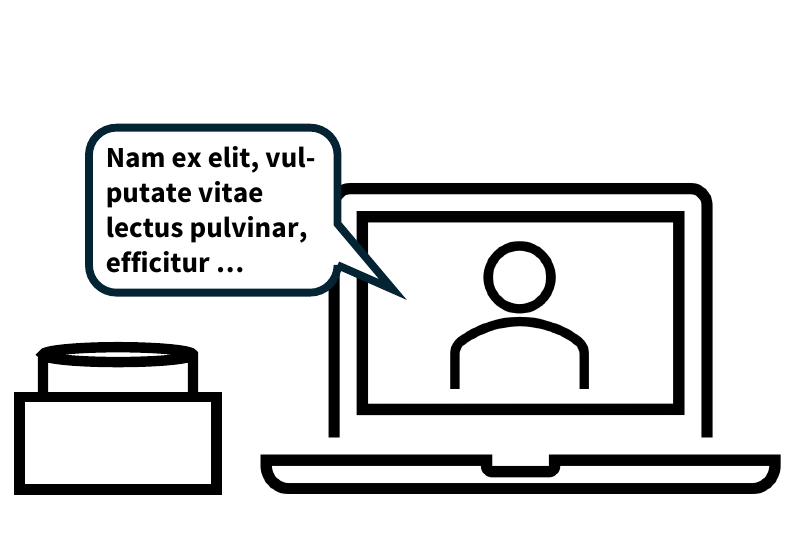}  
  \caption{Presentation continues}
  \label{fig:sub-second}
\end{subfigure}
  \caption[]{The "Huh?" button\protect\footnotemark
  in action: a method to improve  the individual understanding of viewers of video lectures}
  \label{fig:teaser}
\end{figure}

\footnotetext{which is actually a software feature but here depicted as physical device for the purpose of illustration}

Recent advancements in large language models (LLMs) have sparked a revolution in the field of natural language processing, demonstrating remarkable proficiency in generating text that is contextually relevant and coherent~\cite{brown2020}. We propose to employ this technology to augment individual learning and comprehension within the context outlined above. The result is a solution that provides a low hurdle for follow-up when something is not fully understood, as well as rephrased and expanded explanations like in a face-to-face situation. Concretely, we propose to enhance online video with a feature which allows the user to signal the need for clarification at anytime by pressing a button or saying a keyword. The video will stop and the last phrases of the video will get rephrased and explained by a LLM which has access to the transcription of the video. In case the user continues to have comprehension problems, repeating the signal will trigger the system to simplify and expand the explanation further. Figure~\ref{fig:teaser} shows the procedure. For the purpose of illustration, the user interface is depicted as button. In reality, the described approach is a feature which could get integrated in media streaming platforms or in "listening" virtual assistants, and get activated in the software interface or via voice trigger detection.

\begin{figure}[h]
\centering
  \includegraphics[width=0.95\linewidth]{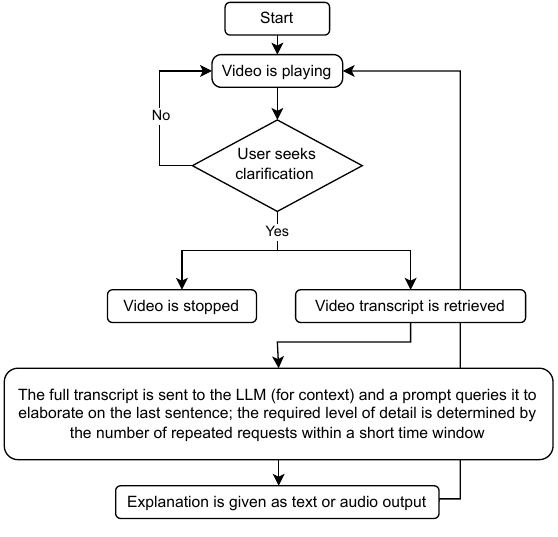}
  \caption{Proposed method for enhancing a recorded video lecture with AI to increase the viewer's individual understanding. The user could trigger the help request by pressing a button or by saying a signal word.}
  \label{fig:flowchart}
\end{figure}

\begin{figure*}[h!]
\begin{subfigure}[t]{.35\textwidth}
    \centering
    \footnotesize
    \begin{tabular}{|m{0.8em}||m{17em}|} 
    \hline 
    \rotatebox[origin=c]{90}{Input (transcript)} & [...] I'd like to build out something like ChatGPT. But we're not going to be able to, of course, reproduce ChatGPT. This is a very serious, production-grade system. It is trained on a good chunk of the internet, and then there's a lot of pre-training and fine-tuning stages to it. So it's very complicated. What I'd like to focus on is just to train a Transformer-based language model, and in our case, it's going to be a character-level language model. I still think that is very educational with respect to how... These systems work, so I don't want to train on the chunk of Internet. We need a smaller data set. In this case, I propose that we work with my favorite toy data set. It's called Tiny Shakespeare. What it is, is basically it's a concatenation of all of the works of Shakespeare, in my understanding. And, so this is all of Shakespeare in a single file. This file is about one     
\\ 
     \hline
     \rotatebox[origin=c]{90}{Output} & The teacher is discussing the complexities of building something akin to ChatGPT, detailing that it is a well-built system trained on a substantial amount of internet data and fine-tuned. Nevertheless, they intend to focus on creating a transformer-based language model, primarily a character-level language model. They reject the idea of training this model on a sizable chunk of the Internet, suggesting a smaller, favorite data set of theirs instead: Tiny Shakespeare, which is a collation of all Shakespeare's works.\\ 
     \hline
    \end{tabular}
    \begin{tikzpicture}[remember picture,overlay]
    \node[anchor=north east,xshift=-0.2cm,yshift=-0.5cm]{\includegraphics[width=1.8cm,height=1.08cm]{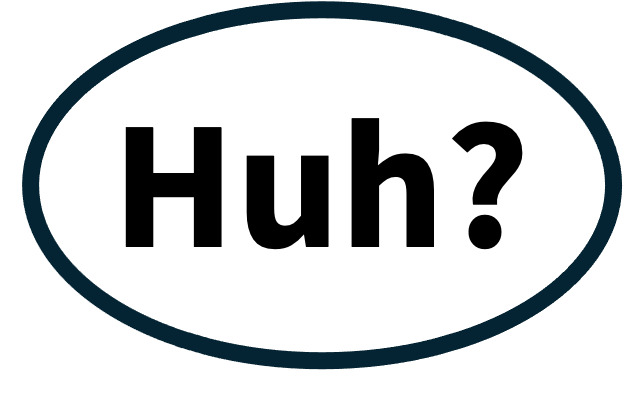}};
    \end{tikzpicture}
    \caption{Computer science}
    \label{fig:lectures-cs}
\end{subfigure}
\begin{subfigure}[t]{.32\textwidth}
    \centering
    \footnotesize
    \begin{tabular}{|m{18em}|} 
    \hline 
     [...] have 'Porterman' analytically. And in homozygous recessive organisms, we will have both releasing alleles genetically. Okay, so basically these different types of organisms, the genotype is telling us what are the different types of genes found in the organisms. The genetic makeup of the organism. So what, is the phenotype? Like I said earlier, phenotype is the physical appearance of an organism or an individual. In heterozygous organisms, capital 'T' and small 't' because here we have the... Dominant allele, so it will mask the expression of recessive allele. So, we will have the tall plant. Okay, you will have tall plants. So, this is the phenotype, tall plant, in case of homozygous dominant where both alleles are dominant, capital T and small T. Here also, we will get tall plant and this tall plant is the phenotype \\ 
     \hline
      In the speaker's conclusion, the term "phenotype" is defined as the actual physical appearance or observable traits of an organism, in this case, a plant. The speaker provides examples of this concept using genetics: In the case of heterozygous organisms, which have one dominant and one recessive allele, the effect of the dominant allele overrides that of the recessive allele, leading to the trait associated with the dominant allele appearing—in this case, tallness in plants. Similarly, in homozygous dominant cases, where both alleles are dominant, the speaker says the plant will also be tall. The observable trait, in this case the height of the plant, is referred to as the phenotype.\\ 
     \hline
    \end{tabular}
    \begin{tikzpicture}[remember picture,overlay]
    \node[anchor=north east,xshift=-0.1cm,yshift=0.5cm]{\includegraphics[width=1.8cm,height=1.08cm]{images/huh.png}};
    \end{tikzpicture}
    \caption{Biology}
\end{subfigure}
\begin{subfigure}[t]{.32\textwidth}
    \centering
    \footnotesize
    \begin{tabular}{|m{18em}|} 
    \hline 
     [...] And we'll go into the reasons why. And we have optimality of competitive equilibria in existence. We have theorems-- existence of competitive equilibria, existence of Nash equilibria. For the Nash equilibrium, we'll go to a model of US financial markets and try to analyze why we see trade fails, where those repo contracts fail, and whether penalties could or should have been in place to prevent those failures. I think I said enough about aggregation and identification already when I went through this the first time. And we end with when the welfare theorems fail, but not stopping there. The analogy with pollution is to fix it by selling rights to pollute. We'll talk about how to handle externalities much more generally. We'll talk about failures of the First Welfare theorem when there's dynamics and an infinite amount of wealth. And likewise, when it fails, we'll see the emergence of money-- both fiat money and Bitcoin. \\ 
     \hline
      The speaker states that they will discuss further failed welfare theorems, using the pollution analogy where the solution has been to sell rights to pollute. They plan to discuss how to manage externalities in a broader context. They also intend to discuss situations where the First Welfare theorem fails due to dynamics and infinite wealth. In such instances, the speaker mentions the emergence of both fiat money and Bitcoin. The speaker believes that it is important to study these issues, and they intend to dive into these topics in depth.\\ 
     \hline
    \end{tabular}
    \begin{tikzpicture}[remember picture,overlay]
    \node[anchor=north east,xshift=-2.1cm,yshift=-0.5cm]{\includegraphics[width=1.8cm,height=1.08cm]{images/huh.png}};
    \end{tikzpicture}
    \caption{Economics}
    \end{subfigure}
\caption{Exemplary results from analyzing three online lectures
: The first row includes the video transcription up to the user's request for further explanation, while the second row has the LLM's additional clarifications.}
\label{exemplary_results}
\end{figure*}

\section{Proof of concept}
The flowchart in Figure~\ref{fig:flowchart} shows the sequential process of our proposal in more detail. To assess the feasibility of our idea, we first decided to carry out a proof-of-concept experiment. For this, we identified three online lectures on YouTube from the fields of computer science, biology and economics\footnote{https://youtu.be/\{kCc8FmEb1nY | wpraShKAJcw | XSTSfCs74bg\}}. We downloaded their transcripts and added missing punctuation using a GPT-4 model~\cite{openai2023gpt4}. To simulate the moment when a user asks for clarification by sending the signal, we truncated the transcript at random points (while allowing for a buffer at the beginning to establish some context). We sent the resulting text snippets to a GPT-4 model along with a prompt which requests to explain the last phrase in detail. We tested several prompts and found that, for example, \textit{"Use a third person singular perspective, referring to the speaker. Take the last sentence of this text which ends with a full stop, and explain it in your own words: [...]"} achieves the desired results. Finally, we tested the setup several times for all three lectures and, despite some serious transcription errors, found the results promising for a proof-of-concept. Exemplary results for each lecture examined can be found in Figure~\ref{exemplary_results}. We also published a Jupyter notebook on GitHub with the source code of the experiment and the transcript files we used for testing\footnote{\url{https://anonymous.4open.science/r/the-huh-button-A371/}}.

Following these encouraging results, we decided to proceed with the implementation of a prototype. To demonstrate the scalability of our idea, we developed a Javascript plugin that seamlessly integrates with YouTube videos. The action is triggered by a button which is rendered below the video player. In order to guarantee low latency and easy technical integration, all explanations have been pre-produced and made accessible as web resources via HTTP. Explanations were created at 5-second intervals and on two different levels: For the first kind of explanations, the LLM was instructed to explain the last sentence based on all information given up to that point. For the second-level explanations, the LLM was asked to consider the last two sentences and thus take a broader perspective in its response. It was also explicitly asked to use simple language. Those clarifications have been included to address situations where the initial explanation may not have been sufficient to the user. In our experiments, both selected scopes produced reasonable results; however, the methodological determination of the optimal text length as basis for the explanations remains a subject for future investigations. 

\begin{figure}[h]
\centering
  \includegraphics[width=1\linewidth]{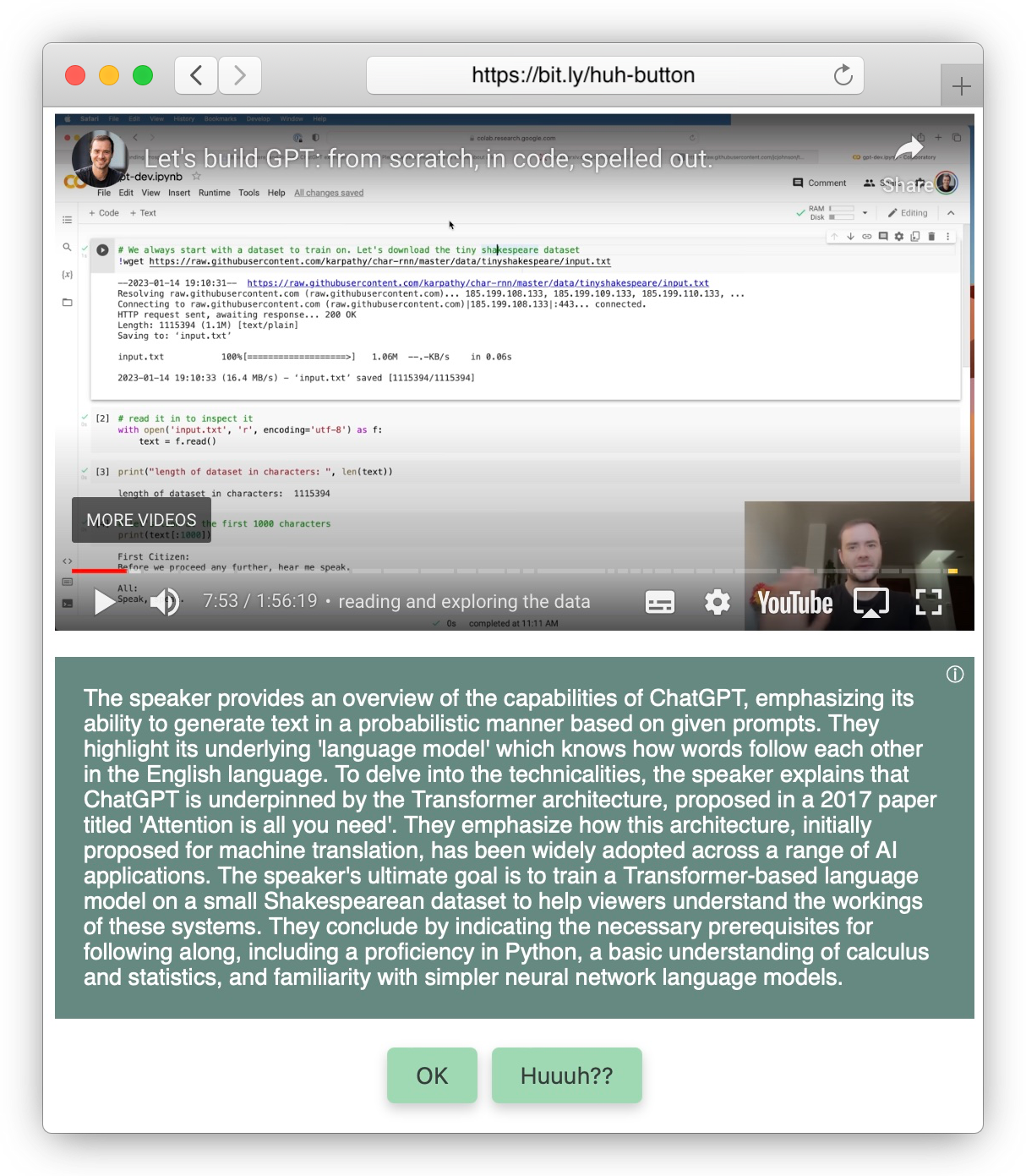}
  \caption{Screenshot of the implemented "Huh?" button, integrated with a YouTube video lecture on a computer science topic~\cite{andrej}. A simplified, second-level explanation is available if the first explanation is not sufficient. Demos have been published online\protect\footnotemark.
  }
  \label{fig:prototype} 
\end{figure}

As querying LLMs is costly, explanations were only generated for subsets of the videos. When no explanations are available, the button assumes an inactive state. As educational video, the computer science lecture from the previous experiment was used (Figure \ref{fig:lectures-cs}). To test the scenario in a language other than English, we also used a lecture by a German philosopher. In the latter case, the poorer quality of the transcriptions posed a challenge but the application's functionality was ultimately not affected. Both demos are available online for testing\footnote{\url{https://github.com/borisruf/the-huh-button}}, a screenshot of the prototype can be found in Figure~\ref{fig:prototype}.

\section{Environmental impact}
It is well known that LLMs consume a considerable amount of energy and therefore have a significant impact on the environment~\cite{luccioni2023power}. For our prototype, we used the Python library "API Emission Tracker"\footnote{\url{https://pypi.org/project/api-emissions-tracker/}} to measure the carbon emissions linked to model inference in the development phase. For the English computer science lecture, explanations were generated by a GPT-4 model for a duration of 13 minutes and 35 seconds (from timestamps 00:10 to 13:45). In total, 390,962 prompt tokens and 37,435 completion tokens were used and the carbon footprint was estimated at 150.7 kg CO2e.
For the German philosophy lecture, we used the same model and generated explanations for a duration of 15 minutes and 35 seconds (from timestamps 00:30 to 16:05). In total, 531,619 prompt tokens and 63,727 completion tokens were used and the estimated carbon footprint amounted to 209.4 kg CO2e. During the production phase, there are no LLM inferences occurring.

Consequently, the presented application shows a rare use case where the utilization of LLMs can be subject to significant caching. This is due to the linear nature of the approach, where the explanation depends only on the timestamp, but is the same for all users. As real-time processing is not required during the development phase, AI model queries can be processed in geographic regions with low carbon factors and scheduled during off-peak times. During operation, interaction is only based on lightweight HTTP requests loading text, which has a negligible carbon footprint compared to using LLM models. This methodology helps reduce the overall carbon footprint during the project life time significantly.

\footnotetext{\url{https://borisruf.github.io/the-huh-button/index-youtube.html}}

\section{Conclusion and future work}
We propose an actionable technology solution in the field of education that could help learners raise the level of understanding until more sophisticated, fully interactive individual tutors become a reality. To achieve this, we suggest to leverage the availability of millions of one videos and lectures with educational content, the accessibility of their transcripts, plus the impressive linguistic abilities of the most recent large language models. The low inhibition threshold when interacting with a computer system and the simplest conceivable user interface of requesting help lead us to expect a high adoption rate. The positive effect of rewording on comprehension, and the possibility of simplified repetition suggest a higher effectiveness than classic "rewinding". Additionally, the presented work offers initial insights into caching approaches for LLM-supported applications through the architectural implementation decisions made.


In a next step, we will confirm the preliminary findings of this working paper by conducting a more rigorous study with human participants to evaluate the usefulness of the provided explanations. Additionally, the accuracy of the output needs to be assessed more systematically, considering the problem of hallucinations of LLMs. Finally, ways to reduce the environmental impact through more dynamic caching approaches should be evaluated.


\bibliographystyle{IEEEtran}
\bibliography{bibliography}

\balance

\end{document}